\documentclass[aps,twocolumn,preprintnumbers,%
superscriptaddress,floatfix]{revtex4}
\usepackage[dvips]{graphics}
\usepackage{amsmath,amsfonts,bm}
\usepackage{epsfig}
\begin{document}
\newcommand{\be}{\begin{eqnarray}}
\newcommand{\ee}{\end{eqnarray}}
\def\lsim{\mathrel{\rlap{\lower3pt\hbox{\hskip1pt$\sim$}}
     \raise1pt\hbox{$<$}}} 
\def\gsim{\mathrel{\rlap{\lower3pt\hbox{\hskip1pt$\sim$}}
     \raise1pt\hbox{$>$}}} 
\def\N{${\cal N}\,\,$}
\def\prl{Phys. Rev. Lett.}
\def\np{Nucl. Phys.}
\def\pr{Phys. Rev.}
\def\pl{Phys. Lett.}
\def\la{\langle}\def\ra{\rangle}
\def\del{\partial}
\def\calL{\cal L}\def\calK{\cal K}
\def\hatn{\hat{n}}\def\Amu{{\cal A}_\mu}\def\A{{\cal A}}
\newcommand\<{\langle}
\renewcommand\>{\rangle}
\renewcommand\d{\partial}
\newcommand\LambdaQCD{\Lambda_{\textrm{QCD}}}
\newcommand\Tr{\mathrm{Tr}\,}
\newcommand\+{\dagger}
\newcommand\g{g_5}
\def\bi{\bibitem}

\newcommand{\msun}{\mbox{$M_\odot$}}
\newcommand{\rsun}{\mbox{$R_\odot$}}

\title{Kaons in Dense Half-Skyrmion Matter}
\author{Byung-Yoon Park}
 \affiliation{\it Department of Physics, Chungnam National University, Daejon 305-764, Korea}
\author{Joon-Il Kim}
 \affiliation{Department of Physics, Florida State University, Tallahassee, FL 32306, USA}
\author{Mannque Rho}
 \affiliation{ Institut de Physique Th\'eorique,  CEA Saclay, 91191 Gif-sur-Yvette C\'edex, France\\
 and Department of Physics, Hanyang University, 133-791 Seoul, Korea}

\begin{abstract}
Dense hadronic matter at low temperature is expected to be in crystal and at high density make a transition to a {\em chirally restored but color-confined} state which is a novel phase hitherto unexplored. This phase transition is predicted in both skyrmion matter in 4D and instanton matter in 5D, the former in the form of half-skyrmions and the latter in the form of half-instantons or dyons. We predict that when $K^-$'s are embedded in this half-skyrmion or half-instanton (dyonic) matter which may be reached not far above the normal density, there arises an enhanced attraction from the  soft dilaton field figuring for the trace anomaly of QCD and the Wess-Zumino term. This attraction may have relevance for a possible strong binding of anti-kaons in dense nuclear matter and for kaon condensation in neutron-star matter. Such kaon property in the half-skyrmion phase is highly non-perturbarive and may not be accessible by low-order chiral perturbation theory. Relevance of the half-skyrmion or dyonic matter to compact stars is discussed.
\end{abstract}

\date{\today}

\newcommand\sect[1]{\emph{#1}---}

\maketitle
\sect{I. The Problem and Results}
There is a compelling indication from baryonic matter simulated on crystal lattice that as density increases beyond the normal nuclear matter density $n_0$, there emerges a phase with vanishing quark condensate symptomatic of chiral symmetry restoration but colored quarks still confined. This has been observed~\cite{half-skyrmions,park-vento} with skyrmions put in an fcc crystal which turn into half-skyrmions in a cc configuration as density reaches $n=n_{1/2} > n_0$ where the quark condensate $\la\bar{q}q\ra$ goes to zero but the pion decay constant $f_\pi$ remains non-zero, implying that hadrons are relevant degrees of freedom there although chiral symmetry is restored. This has also been shown~\cite{dyons} to take place with instantons in 5D that figure in holographic dual QCD (hQCD) placed in an fcc crystal that split into half-instantons in the form of dyons in a bcc crystal configuration. While the skyrmion matter is constructed either with the pion field only or with the pion field plus the lowest lying vector mesons $\rho$ and $\omega$ that we shall denote $\tilde{\rho}$, the instanton matter arises as a solitonic matter in 5D which when viewed in 4D, contains an infinite tower of both vector and axial vector mesons. A single baryon is found to be much better described as an instanton in hQCD~\cite{instanton-baryon} which is justified for both large $N_c$ and large 't Hooft constant $\lambda=g_{YM}^2 N_c$ than the skyrmion baryon in large $N_c$ QCD. This is particularly so for quantities captured in quenched lattice QCD calculations. Now a highly pertinent question is: How does a meson behave in the medium consisting of a large number of these solitons?

This issue was addressed in \cite{half-skyrmions} for fluctuations of pions in dense medium. What one learned from that calculation was limited because the pion, being nearly a genuine Goldstone boson, is largely protected by chiral symmetry and singling out medium effect would require a high accuracy in theory -- such as, e.g., high-order $1/N_c$ corrections -- which the skyrmion description is not capable of providing. The story however is quite different with the kaons. In fact, the Callan-Klebanov model~\cite{CK} that describes the hyperons as bound states of the fluctuating $K^-$'s with an $SU(2)$ soliton has been amazingly successful as recently reviewed in \cite{scoccola}. Here the Wess-Zumino term that encodes chiral anomalies plays a singularly important role in providing the binding of anti-kaons to an $SU(2)$ soliton. A similar observation was made for the two-baryon system $K^- pp$ in \cite{Nishikawa-Kondo} where it was found that there is a substantial increase in attraction between the kaon and the nucleons when the latter interact at a shorter distance.

In this article, we consider negatively charged kaons fluctuating in the medium described as a dense solitonic background. This involves two very important issues in physics of dense hadronic matter. One is that there is a possibility that $K^-$ can trigger strongly correlated mechanism to compress hadronic matter to high density~\cite{yamazaki}. Such a mechanism is thus far unavailable in the literature and may very well be inaccessible in perturbation theory~\cite{vonnegut}. The other issue is the role of kaon condensation in dense compact star matter which has ramifications on the minimum mass of black holes in the Universe and cosmological natural selection~\cite{BLR-CNS}. It would be most appealing and of great theoretical interest to address this problem in terms of the instanton matter given in hQCD, which has the potential to also account for shorter-distance degrees of freedom via an infinite tower of vector and axial vector mesons. However numerical work in this framework is unavailable. A skyrmion matter consisting of pion and $\tilde{\rho}$ has been studied~\cite{PRV} but has not yet been fully worked out. We shall therefore take the simple Skyrme model implemented with two key ingredients, viz, the Wess-Zumino term and the ``soft dilaton" field $\chi_s$ that accounts for scale symmetry tied to spontaneously broken chiral symmetry as precisely stated in ~\cite{LR-dilatons}. Our task is to understand kaon fluctuations in the background given by the skyrmion matter described by this Lagrangian. What makes this approach different from others is that it exploits the close connection between scale symmetry breaking encoded in the dilaton condensate which can be restored \`a la Freund-Nambu~\cite{Freund-Nambu,LR-dilatons} and chiral symmetry breaking encoded in the skyrmion matter, both linked to the mass of light-quark hadrons. There are certain fine-tunings required with the parameters of the model to achieve a quantitative comparison with nature, which we will eschew in this article. We shall instead focus more on robust qualitative features. The basic approximation that we will adopt -- which will have to be ultimately justified -- is that the back-reaction of kaon fluctuations on the background matter can be ignored, which is consistent with large $N_c$ consideration.

\begin{figure}  
\centerline{\epsfig{file=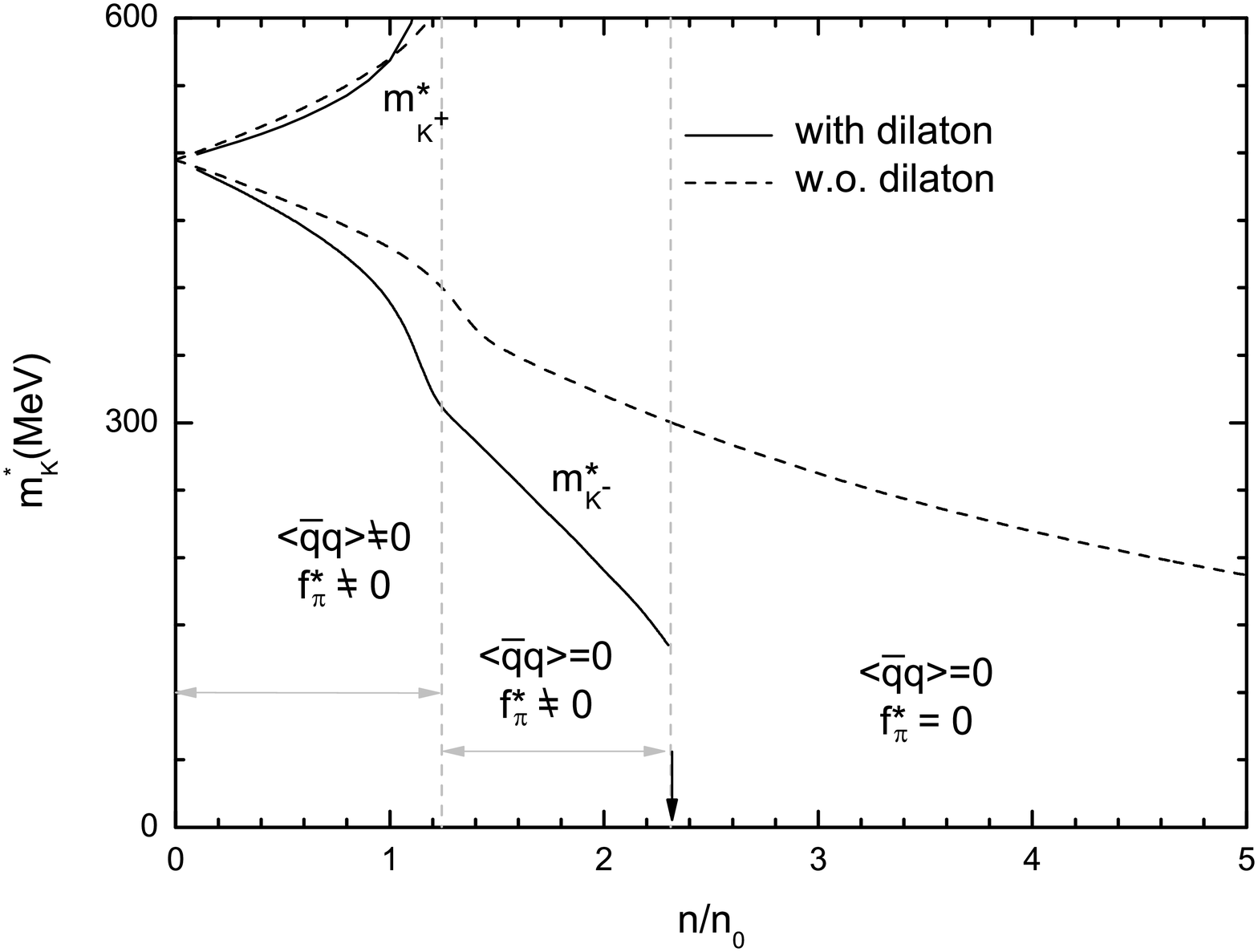, width=8cm, angle=0}}
\centerline{\epsfig{file=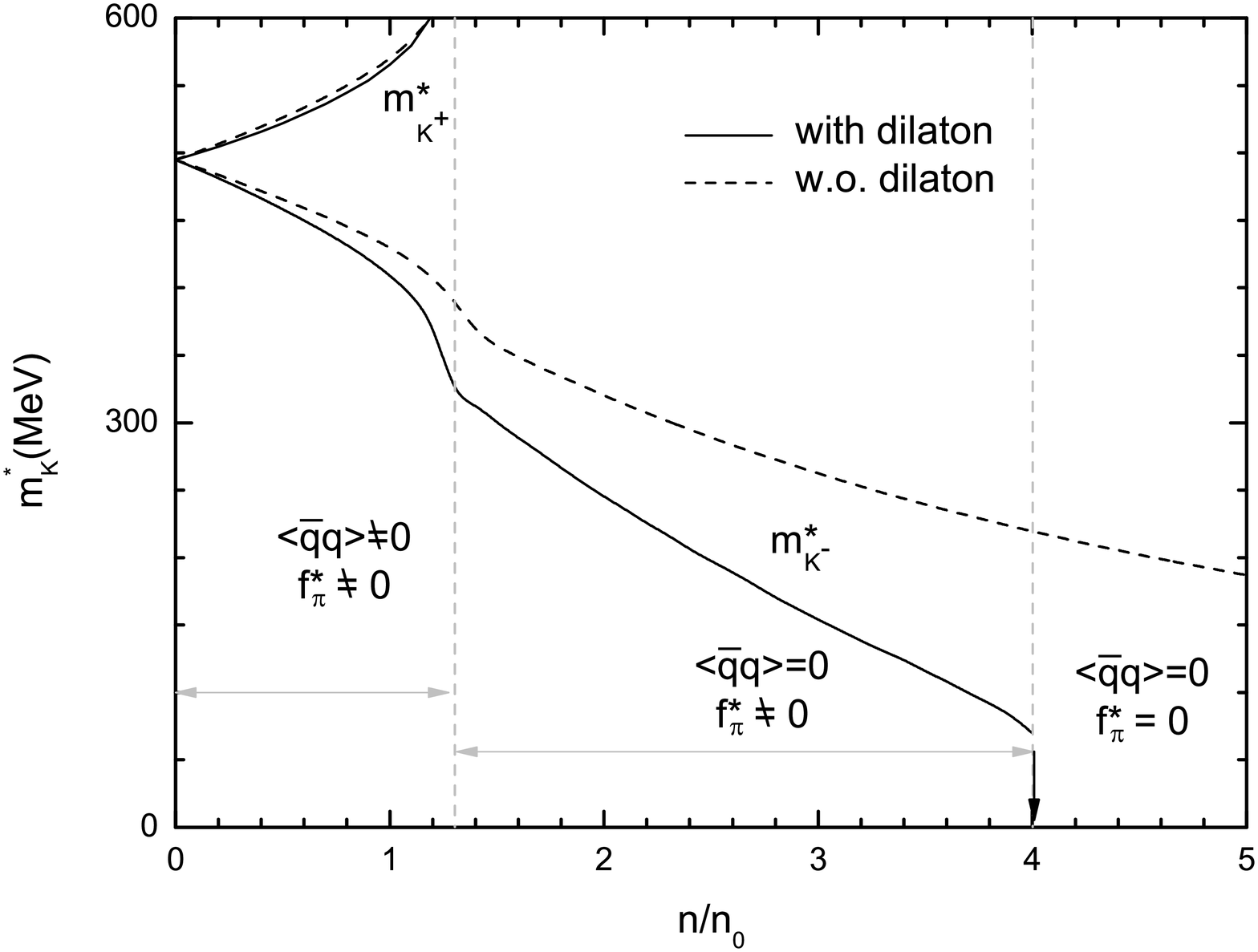, width=8cm, angle=0}}
\caption{ $m^*_{K^\pm}$ vs. $n/n_0$ (where $n_0\simeq 0.16$ fm$^{-3}$ is the normal nuclear matter density) in dense skyrmion matter which consists of three phases: (a) $\la\bar{q}q\ra\neq 0$ and $f_\pi^*\neq 0$, (b) $\la\bar{q}q\ra=0$ and $f_\pi^*\neq 0$ and (c) $\la\bar{q}q\ra=0$ and $f_\pi^*=0$. The parameters are fixed at $\sqrt{2}ef_\pi=m_\rho=780$ MeV and dilaton mass $m_\chi=600$ MeV (upper panel) and $m_\chi=720$ MeV (lower pannel). }
\label{fig1}
\end{figure}

Our result is summarized in Fig.~\ref{fig1}.

We find in the model three different regimes in properties of the kaon as density increases. At low density with chiral symmetry spontaneously broken with  $\la\bar{q}q\ra\propto {\rm Tr} U\neq 0$ and $f_\pi\neq 0$, the state is populated by skyrmions and the mass of the kaon propagating therein drops at the rate controlled by chiral perturbation theory valid at low density. This behavior continues up to the density $n_{1/2}$ at which the skyrmion matter turns into a half-skyrmion matter characterized by ${\rm Tr}U=0$ so chiral symmetry is restored but $f_\pi\neq 0$. In this phase, the kaon mass undergoes a much more dramatic decrease until it vanishes -- in the chiral limit -- at the critical density $n_c$ at which ${\rm Tr} U=f_\pi=0$. The region between $n_{1/2}$ and $n_c$,  dubbed as ``hadronic freedom" regime~\cite{LR-dilatons} -- which also played an important role in explaining dilepton processes in heavy-ion collisions~\cite{BHHRS}, is most likely inaccessible by low-order chiral perturbation theory. Given the extreme truncation of the model used here, it makes little sense to attempt to pin down precisely the onset density of the half-skyrmion phase. The best guess would be that it figures at a density between 1.3 and 3 times the normal nuclear density $\approx 0.16$ fm$^{-3}$, the range indicated in Fig.~\ref{fig1} for the set of parameters chosen.

\sect{II. The Lagrangian}
To construct dense nuclear matter into which kaons are to be embedded, we take the $SU(2)_f$ Skyrme Lagrangian given in terms of the pion field only~\cite{skyrme} and construct a dense baryonic matter by putting the skyrmions on a crystal. This Skyrme Lagrangian can be considered as an effective low-energy Lagrangian valid at large $N_c$ in which all other degrees of freedom are integrated out. In fact, it is seen in hQCD that the Skyrme quartic term -- considered in the past as {\em ad hoc} -- does arise naturally and  uniquely, capturing physics of shorter-distance than that of the lowest vector excitations, i.e., $\tilde{\rho}$. There is however one crucial ingredient that needs to be implemented to the Skyrme Lagrangian, namely, the dilaton field that accounts for the trace anomaly of QCD. In considering dense matter, it is essential that the scale-symmetry breaking encoded in the trace anomaly be mapped to the spontaneously breaking of chiral symmetry. This point was implicit already in the 1991 proposal for scaling of hadron masses~\cite{BR91} but it was in \cite{LR-dilatons} that the dilaton field that figures in the connection was clearly identified. There is a subtlety in distinguishing the ``soft dilaton" $\chi_s$ with which we are concerned here and the ``hard dilaton" $\chi_h$ which is associated with the asymptotically free running of the color gauge coupling constant reflecting scale symmetry breaking in QCD. This is discussed in \cite{LR-dilatons}. What is important for our purpose is that the condensate of the soft dilaton vanishes across the chiral restoration point whereas that of the hard dilaton remains non-zero across the critical point. The vanishing of the soft dilaton condensate is directly connected to the vanishing of the quark condensate (in the chiral limit) and hence to chiral symmetry as shown in the case of dileptons in heavy-ion collisions~\cite{BHHRS}. It has not yet been rigorously established but is plausible~\cite{BHHRS,HKR:VM} that the same holds in the case of density. In this article, we will simply assume that it does and deal uniquely with the soft component which we will denote simply by $\chi$.

Extended to three flavors and implemented with the dilaton field $\chi$, the Skyrme Lagrangian we shall use takes the form~\cite{PRV,LR-dilatons}
\begin{eqnarray}
{\cal L}_{sk}
&=& \frac{f^2}{4} \left(\frac{\chi}{f_\chi}\right)^2
{\rm Tr} (L_\mu L^\mu) + \frac{1}{32e^2}{\rm Tr} [L_\mu, L_\nu]^2\nonumber\\
&& +\frac{f^2}{4}\left(\frac{\chi}{f_\chi}\right)^3
{\rm Tr}{\cal M} (U+U^\dagger-2)
\nonumber \\
&&+\frac{1}{2}\partial_\mu \chi\partial^\mu \chi + V(\chi)\label{SK}
\label{lag1}
\end{eqnarray}
where $V(\chi)$ is the potential that encodes the trace anomaly involving the soft dilaton, $ L_{\mu} = U^{\dagger} \partial_{\mu} U $, with
$U$ the chiral field taking values in $SU(3)$ and $f_\chi$ is the vev of $\chi$. We shall ignore the pion mass for simplicity, so the explicit chiral symmetry-breaking mass term is given by the mass matrix ${\cal M} = diag \left( 0, 0, 2m_{K}^2\right)$. In $SU(3)_f$, the anomaly term, i.e., the Wess-Zumino term, 
${\cal S}_{WZ} = - \frac{i N_C}{240 \pi^2} \int d^5 x
\varepsilon^{\mu\nu\lambda\rho\sigma}
\mbox{Tr} \left( L_{\mu} L_{\nu} L_{\lambda} L_{\rho} L_{\sigma} \right)$,
turns out to play a crucial role in our approach:

\sect{Fluctuating Kaons in the Skyrmion Matter}
In close analogy to the Callan-Klebanov scheme~\cite{CK}, we consider the fluctuation of kaons in the background of the skyrmion matter $u_0$, following \cite{riskaetal}, as
\begin{equation}
  U(\vec x,t) = \sqrt{U_K(\vec x,t)} U_0(\vec x) \sqrt{U_K(\vec x,t)},\label{CKansatz}
\end{equation}
\begin{equation}
  U_K(\vec x,t) =
  e^{\frac{i}{\sqrt{2}f_{\pi}}
\left( \begin{array}{cc}
0 & K \\ K^\dagger &
0 \end{array} \right)},\
  U_0(\vec x)
    = \left( \begin{array}{cc} u_0(\vec x) & 0 \\ 0 & 1 \end{array} \right).
\end{equation}
Substituting (\ref{CKansatz}) into (\ref{lag1}) and the Wess-Zumino term,
we get the kaon Lagrangian in the background matter field $u_0(x)$
\begin{eqnarray}
{\cal L}_{K}
&=&
\bigg(\frac{\chi_0}{f_\chi} \bigg)^2 \dot K^{\dagger}G\dot K
- \bigg(\frac{\chi_0}{f_\chi} \bigg)^2 \partial_iK^{\dagger}G\partial_iK\nonumber\\
&& \hskip 2em - \bigg(\frac{\chi_0}{f_\chi} \bigg)^3 m_{K}^2 K^{\dagger} K
\nonumber\\
&& \hskip 2em
 + \frac{1}{4} \bigg(\frac{\chi_0}{f_\chi} \bigg)^2
   \left( \partial_\mu K^{\dagger} V^{\mu}(\vec x) K
 - K^{\dagger} V_\mu(\vec x) \partial^{\mu}K \right),
\nonumber \\
&&  \hskip 2em
+ \frac{iN_c}{4f_{\pi}^2} B^0
 \left( K^{\dagger}G\dot K - \dot K^{\dagger} GK \right).
\end{eqnarray}
where $\chi_0(\vec{x})$ is the classical dilaton field and
\begin{eqnarray}
V_\mu(\vec{x})
&=& \frac{i}{2} [(\partial_\mu u_0^\dagger)u_0 - (\partial_\mu u_0) u_0^\dagger],
\label{V} \\
G(\vec x) &=& \frac{1}{4} (u_0+u_0^{\dagger}+2),
\\
B^{\mu}(\vec x) &=& \frac{1}{24\pi^2} \varepsilon^{\mu\nu\lambda\sigma}
 {\rm Tr}\left( u_0^{\dagger} \partial_{\nu} u_0
 u_0^{\dagger} \partial_{\lambda} u_0 u_0^{\dagger}
 \partial_{\sigma} u_0 \right).
\end{eqnarray}
In the spirit of mean field approximation, we will take the space average on the
background matter fields $u_0$ and $\chi_0$ and obtain
\begin{equation}
{\cal L}_{K}
= \alpha ( \partial_\mu K^{\dagger} \partial^\mu K)
+ i \beta ( K^{\dagger} \dot K - \dot K^{\dagger} K )
- \gamma K^{\dagger} K
\label{MFA}
\end{equation}
where
\begin{eqnarray}
\alpha = \left< \kappa^2 G \right>,
\beta = \frac{N_c}{4f_{\pi}^2} \left< B^0 G \right>,
\gamma = \left<\kappa^3 G \right> m_{K}^2
\end{eqnarray}
with $\kappa=\chi_0/f_\chi$.
Lagrangian (\ref{MFA}) yields a dispersion relation for the kaon in the
skyrmion matter as
\begin{equation}
\alpha( \omega_K^2 - p_K^2) + 2 \beta \omega_K + \gamma = 0.
\end{equation}
Solving this for $\omega_K$ and taking the limit of $p_K \rightarrow 0$,
we have
\begin{equation}
m_K^* \equiv \lim_{p_K \rightarrow 0} \omega_K
= \frac{-\beta + \sqrt{\beta^2 + \alpha\gamma}}{\alpha}.
\end{equation}
This equation will be used for evaluating the in-medium effective kaon mass.

\sect{III. Skyrmion Crystal}
The key element in Eq.~(\ref{MFA}) for the kaonic fluctuation is the backfround $u_0$ which reflects the ``vacuum" modified by the dense skyrmion matter. The classical dilaton field tracks the quark condensate affected by this skyrmion background $u_0$ that carries information on chiral symmetry of dense medium. Our approach here is to describe this background $u_0$ in terms of crystal configuration. The pertinent $u_0$ has been worked out in detail in \cite{half-skyrmions}, from which we shall simply import the results for this work. As shown there, skyrmions put on an fcc -- which is the favored crystal configuration -- make a phase transition at a density $n_{1/2}$ to a matter consisting of half-skrymions. With the parameters of the Lagrangian picked for the model, we find $n_{1/2}\sim 1.3 n_0$ but this is quite uncertain. As noted, our guess would range from slightly above $n_0$ to $\sim 3 n_0$.  This is also where the instanton-to-half-instantons (or dyons) transition takes place in hQCD~\cite{dyons}. In this paper, we will not attempt a quantitative estimate but consider the above density as a ball-park value.

\sect{IV. Vector Mode}
An important aspect of the half-skyrmion state is that the quark condensate $\la\bar{q}q\ra$ -- which is proportional to ${\rm Tr}(u_0+u_0^\dagger)$ -- is zero in this phase but the pion decay constant $f_\pi$ -- which is proportional to $\la\chi_0\ra$ in medium -- could be non-zero. This means that the half-skyrmions are hadrons, not deconfined quarks. There is a resemblance to fractionized electrons in (2+1) dimensions in condensed matter physics~\cite{senthil}. In hidden local symmetry theory~\cite{HLS}, that $\la\bar{q}q\ra=0$ {\em and} $f_\pi\neq 0$ implies that $f_\pi=f_s$ with $\la 0|A_\mu|\pi\ra=ip_\mu f_\pi$ and $\la 0|V_\mu|s\ra=ip_\mu f_s$ where $s$ is the longitudinal component of the $\rho$ meson.  This corresponds to the ``vector mode" conjectured by Georgi~\cite{georgi} to be realized in the large $N_c$ limit of QCD. Such a mode does not exist in QCD proper with Lorentz invariance. However one can think of it as an emergent symmetry in the presence of medium, which dense matter provides. The hadronic freedom regime mentioned above encompasses this mode with $a\equiv f_s/f_\pi\approx 1$ and $g\approx 0$ (where $g$ is the hidden gauge coupling constant) near the critical point ($T_c$ or $n_c$).

\sect{V. Effective Kaon Mass}
We now discuss the results given in Fig.~\ref{fig1} in some detail.

The results are given for two values of dilaton mass $m_\chi=600$ MeV and 720 MeV. As explained in \cite{LR-dilatons}, the dilaton mass is known neither experimentally nor theoretically. We have taken two values which we consider reasonable for the ``soft dilaton." The former corresponds to the lowest scalar excitation seen in nature and the latter to the effective scalar needed in Fermi-liquid description of nuclear matter implementing BR scaling~\cite{song}. As shown in \cite{half-skyrmions}, the density at which the half-skyrmion matter appears is independent of the dilaton mass. In fact, the $n_{1/2}$ is entirely dictated by the parameters that give the background $u_0$. For the skyrmion background, we pick $\sqrt{2}ef_\pi$ -- which is the vacuum $\rho$ mass at tree order-- to be $\sim 780$ MeV which fixes $e$ for the given $f_\pi\approx 93$ MeV~\cite{footnote1}. What is controlled by the dilaton mass are the density $n_c$ at which the quark condensate vanishes together with the pion decay constant and more significantly, the rate at which the kaon mass drops as density exceeds $n_0$.

What is novel in our model is that the behavior of the kaon is characterized by three different phases, not two as in conventional approaches. In the low density regime up to $\sim 1.3n_0$, the kaon interaction can be described by standard chiral symmetry treatments. For instance, the binding energy of the kaon at nuclear matter density comes out to be $\sim 110 (\sim 80)$ MeV for $m_\chi= 600 (720)$ MeV. This is what one would expect in chiral perturbation theory (see \cite{gal} for review). Going above $n_0$, however, as the matter enters the half-skyrmion phase with vanishing quark condensate and non-zero pion decay constant, the kaon mass starts dropping more steeply. This property is consistent with what is observed in the ``hadronic freedom" regime in the approach to kaon condensation that starts from the vector manifestation fixed point of hidden local symmetry~\cite{BLPR}. This form of matter -- which is undoubtedly highly nonperturbative -- is most likely unamenable to a chiral perturbation approach. Finally at $n_c$ at which both the quark condensate and the pion decay constant vanish, the kaon mass vanishes. This happens at $n_c\approx 2.3 (4.0) n_0$ for the dilaton mass $m_\chi=600 (720)$ MeV. The reason for this is that the dilaton condensate, $\la\chi\ra$, vanishes with the restoration of soft scale symmetry \`a la Freund-Nambu explained in \cite{LR-dilatons}. This means that ``kaon condensation" in symmetric nuclear matter takes place at the point at which the scale symmetry associated with the soft dilaton is restored. It is not clear whether this takes place before, or coincides with, deconfinement which requires the intervention of the ``hard" component of the gluon condensate ignored in \cite{LR-dilatons}.

\sect{VI. Compact-Star Matter and Kaon Condensation} Thus far,  the kaon is treated at the semiclassical level as a quasiparticle bound to the skyrmion matter. To understand what this represents in nature, we should note that when one quantizes the system where a kaon is bound to a single skyrmion, the bound system gives rise to the hyperons, $\Lambda$ and $\Sigma$, as shown by Callan and Klebanov~\cite{CK}. This would suggest that kaons bound to a skyrmion matter, when collective-quantized, would correspond to hyperonic matter. On the other hand, one can interpret the system in Ginzburg-Landau mean-field theory focusing on the kaon fluctuation and identify the vanishing of the effective kaon mass in medium, $m_K^\star$, as the  signal for kaon condensation. Thus in this picture, {kaon condensation and hyperon condensation are physically equivalent}.

Now a physically interesting question is what does the half-skyrmion matter do to compact stars? What we have done above is to subject the kaon to the skyrmion background which is given in the large $N_c$ limit. In the large $N_c$ limit, there is no distinction between symmetric matter and asymmetric matter. Compact stars have neutron excess which typically engenders repulsion at densities above $n_0$, and hence the asymmetry effect needs to be taken into account. This effect will arise when the system is collective-quantized which gives rise to the leading $1/N_c$ correction to the energy of the bound system. Most of this correction could be translated into a correction to the effective mass of the kaon. We have no estimate of this correction but it is unlikely to be big. What is more important is the energy that arises from neutron excess, called ``symmetry energy" in nuclear physics, labeled $S(n)$ in the energy per particle $E$ of the neutron-rich system appropriate for compact stars,
\begin{eqnarray}
E(n, x)=E(n,x=0) + E_{sym} (n,x)
\end{eqnarray} with
\begin{eqnarray}
E_{sym} (n, x)=S(n) x^2+\cdots
\end{eqnarray} where $x=(P-N)/(N+P)$ with $N(P)$ standing for the neutron(proton) number. The ellipsis stands for higher order terms in $|x| \leq 1$.

In compact-star matter, kaons will condense at a density below the critical density $n_c$ because of the electron chemical potential which tends to increase as the matter density increases~\cite{BTKR}. Now the electron chemical potential is known to be entirely controlled by the symmetry energy $S$ in charge-neutral beta equilibrium systems~\cite{KL}. Thus collective-quantization is required to address at what density kaons will condense. Quantizing the moduli space of multi-skyrmion systems (or multi-instanton systems in hQCD) has not yet been worked out fully.  As mentioned above, the effect of hyperons on electron chemical potential will be automatically taken into account once the moduli quantization is made. Even in the absence of detailed computations, however, we can say at least within the framework of our model that kaons will condense at a relatively low density, say, $n_K \lsim 4n_0$.

\sect{VII. Conclusion} Hidden local symmetry with vector manifestation~\cite{HLS} combined with a soft dilaton accounting for scale symmetry restoration \`a la Freund-Nambu~\cite{LR-dilatons} implies that the anti-kaon mass in dense medium falls more rapidly in the half-skyrmion phase than in the skyrmion phase in which standard chiral perturbation approach should be applicable. It is proposed that this could be relevant to an ``Ice-9" phenomenon in kaon-nuclear systems and kaon condensation in compact star matter. It is also argued that the half-skyrmion phase corresponds to the ``hadronic freedom" regime in density in parallel to that in temperature that figures in dilepton production in heavy-ion collisions. In this paper, the issue was addressed in terms of half-skyrmions in HLS. It would be interesting to analyze the same in terms of half-instantons (or dyons) in hQCD. We hope to address this problem in a future publication~\cite{WCU}.

\sect{Acknowledgments} This work was supported by the WCU project of Korean Ministry of Education, Science and Technology (R33-2008-000-10087-0).

\end{document}